\documentclass{elsart}

\usepackage{epsfig, amssymb}
\newcommand{\fr}{\frac}
\newcommand{\lan}{\langle}
\newcommand{\ran}{\rangle}

\newcommand{\eps}{\varepsilon}
\newcommand{\beq}{\begin{equation}}
\newcommand{\eeq}{\end{equation}}

\begin{document}

\begin{frontmatter}

\title{Periodic orbit theory in fractal drums}

\author{Stefanie Russ$^1$}
\ead{stefanie.russ@physik.uni-giessen.de}
\author{Jesper Mellenthin$^{1,2}$}
\ead{jesper.mellenthin@polytechnique.fr}

\address{$^1$Institut f\"ur Theoretische Physik III, Universit\"at Giessen, \\ D-35392 Giessen, Germany}
\address{$^2$Laboratoire de Physique de la Mati\`ere Condens\'ee, Ecole Polytechnique, F-91128 Palaiseau, France} 

\begin{abstract} 
The level statistics of pseudointegrable fractal drums is studied numerically using 
periodic orbit theory. We find that the spectral rigidity $\Delta_3(L)$, which is a measure for the
correlations between the eigenvalues, decreases to quite small values (as compared to systems
with only small boundary roughness), thereby approaching the behavior of chaotic systems.
The periodic orbit results are in good agreement with direct calculations of $\Delta_3(L)$ 
from the eigenvalues.
\end{abstract}

\end{frontmatter}

\section{Introduction}
Many systems with irregular geometries exist in nature and their physical 
properties (e.g. their electronic and vibrational behavior) have been 
widely investigated in the past.
One class of irregular systems are those, which are formed by an ordered material,
but possess an irregular shape of the boundary. A prominent example are the
fractal drums \cite{sapfrac}, which are constructed by applying the so-called fractal
generator (see Fig.~\ref{bi:geo}(a)) $\nu$ times to a regular square or rectangle. 
Already for prefractal shapes as e.g. for $\nu=1-3$, the eigenstates possess 
energy spectra and localization properties that are well distinct from
systems with smooth boundaries. This has been demonstrated by
numerical simulations \cite{sapgo,PRE,PRL} as well as by experiments on
liquid crystal films \cite{PRL} and acoustic cavities \cite{acoust}.
In this paper, we address the question how the
specific geometry of the systems influence the behavior of the energy spectra
within periodic-orbit theory, which provides a link
between the classical and the quantum mechanical behavior of systems with the same 
shape.

Two universality classes exist with a different classical dynamics,
the chaotic and the regular systems. 
Fractal drums belong to the intermediate class of pseudointegrable 
systems, which are polygons with only rational 
angles $n_i\pi/m_i$, with $n_i, m_i\in \mathbb{N}$ and at least one $n_i>1$
(see e.g. \cite{richens,cheon89,shudo94,Hlushchuk03,biswas1,biswas2,biswas3,mellenthin04} 
and refs. therein). 
They are characterized by their genus number
$g$, which is equal to $1$ for integrable systems and $\infty$ for chaotic systems.
For pseudointegrable systems, $1<g<\infty$. In the specific case of systems with
only right angles, $g=1+N_{3\pi/2}$, where $N_{3\pi/2}$ is the number of
angles ${3\pi/2}$. The different universality classes persist when the systems are 
considered as quantum mechanical potential wells. In this case, the 
statistics of the eigenvalues shows characteristic features
that correspond to the different dynamics of the classical systems.

\section{Spectral Statistics}

The first way to calculate the spectral statistics starts with  
the distribution $P(s)$ of the normalized distances $s_i=
(\eps_{i+1}-\eps_i)/\lan s \ran$ between
two consecutive eigenvalues $\eps_{i+1}$ and $\eps_i$ with the mean distance 
$\lan s \ran$. For integrable systems, $P(s)$ 
follows the Poisson distribution, whereas the $s_i$ of chaotic
systems are Wigner-distributed \cite{Mehta,berry1}.
Here, we consider the spectral rigidity $\Delta_3(L)$ \cite{dyson}, which describes 
the mean correlations in a normalized energy interval of length $L$. We start from the 
integrated density of states $N(\eps)$ of the normalized (''unfolded'') $\epsilon$
where $\lan s\ran=1$, a staircase function that can be approximated by a straight line.
$\Delta_3(L)$ is defined as the least square deviation between $N(\eps)$ and
its best linear fit $r_1-r_2\eps$ 
\begin{equation}\label{delta3}
\Delta_3(L) = \left\lan\int_{E_0-L/2}^{E_0+L/2}
[N(\eps)-r_1-r_2\eps]^2 d\eps
\right\ran,
\end{equation}
in average over many midpoints $E_0$
from an interval $E_0 \in \Delta E=[0.5,1.5]$. 
The constants have been set to $\hbar=2m=a=1$, where $a$ is the lattice
constant and $m$ the mass, leading to an upper band edge $E_{\rm{max}}=8$. 
The range of $E_0$ corresponds to wavelengths being $5-9 a$, much
smaller than the size of the smallest boundary ''teeth'' that scatter around $16a$ for 
the $\nu=2$-drum and $60 a$ for the $\nu=1$-drum. $\Delta E$
contains $3000-5000$ values of $\epsilon$, whereas the integration range $L$
contains only up to $500$ values (see below).
The limiting curves for not too large $L$ are $\Delta_3(L)=L/15$ for integrable systems 
and $\Delta_3(L)=\ln(L)/\pi^2-0.07/\pi^2 + O(L^{-1})$ for the ensemble of Gaussian 
orthogonal matrices (GOE) \cite{Mehta,berry1}, which serves as a good
generally accepted limit for chaotic systems. For large $L$, $\Delta_3$ reaches
a plateau that depends on the small orbits.

The spectral rigidity $\Delta_3(L)$ can also be calculated by periodic orbit theory
(see Fig.~\ref{bi:geo} for some examples of periodic orbits).
In pseudointegrable systems, all periodic orbits form families of equal lengths and 
stay stable when their starting points are shifted to at least one direction along the 
boundary. The simplest families are the 
''neutral orbits'' that bounce between two parallel walls.
Using semiclassical quantum mechanics and neglecting additional 
contributions coming e.g. from diffractive orbits (that start and end at a salient corner), 
$\Delta_3(L)$ under Neumann boundary conditions is given by \cite{biswas1,biswas2,berry1},
\begin{equation}
\label{biswaseq}
\Delta_3(L) =  \left<\frac {\sqrt{E_0}}{4 \pi^3} \sum_{i,j} \frac
{a_ia_j}{\left(\ell_i\ell_j\right)^{3/2}} \cos\left[\sqrt{E_0}(\ell_i-\ell_j)\right] H_{ij}\right>.
\end{equation}
The double sum is carried out over all orbit families
(including repetitions, but only in forward direction) of lengths $\ell_i$ and $\ell_j$. 
$a_i$ and $a_j$ are the areas in phase space that are occupied by the respective orbit 
families and $H_{ij} = F(y_i-y_j)-F(y_i)F(y_j)-
3(\partial F(y_i)/\partial y_i)(\partial F(y_j)/\partial y_j)$, where 
$F(y)=(\sin y)/y$. The argument $L$ enters via $y_i=(L\ell_i A)/(16\pi\sqrt{E_0})$ with 
the system area $A$. It has been shown in \cite{mellenthin04} that for 
pseudointegrable systems, 
the double sum can be restricted to the diagonal terms of $i=j$ and very long orbits can be
neglected. We therefore need the $\ell_i$ and $a_i$ up to some maximum length.
\unitlength 1.82mm
\vspace*{0mm}
\hspace*{5mm}{
\begin{figure}
\begin{picture}(80,45)
\def\epsfsize#1#2{0.6#1}
\put(40,25){\epsfbox{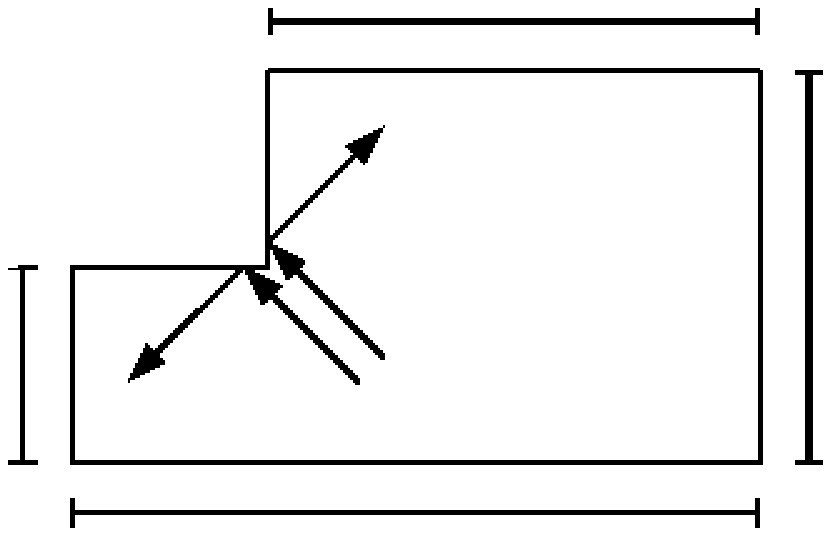}}
\def\epsfsize#1#2{0.35#1}
\put(40,0){\epsfbox{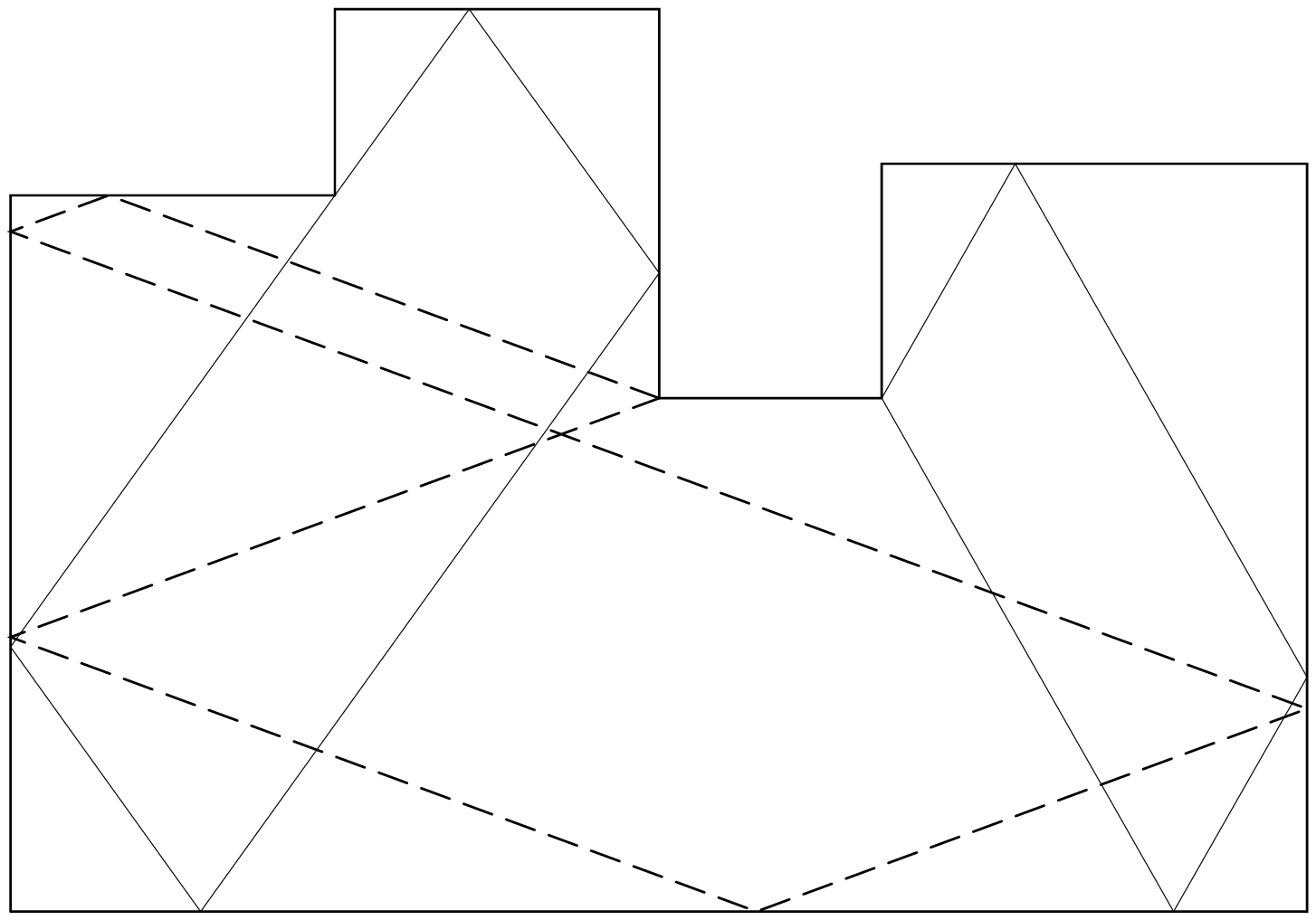}}
\put(0,6){\epsfbox{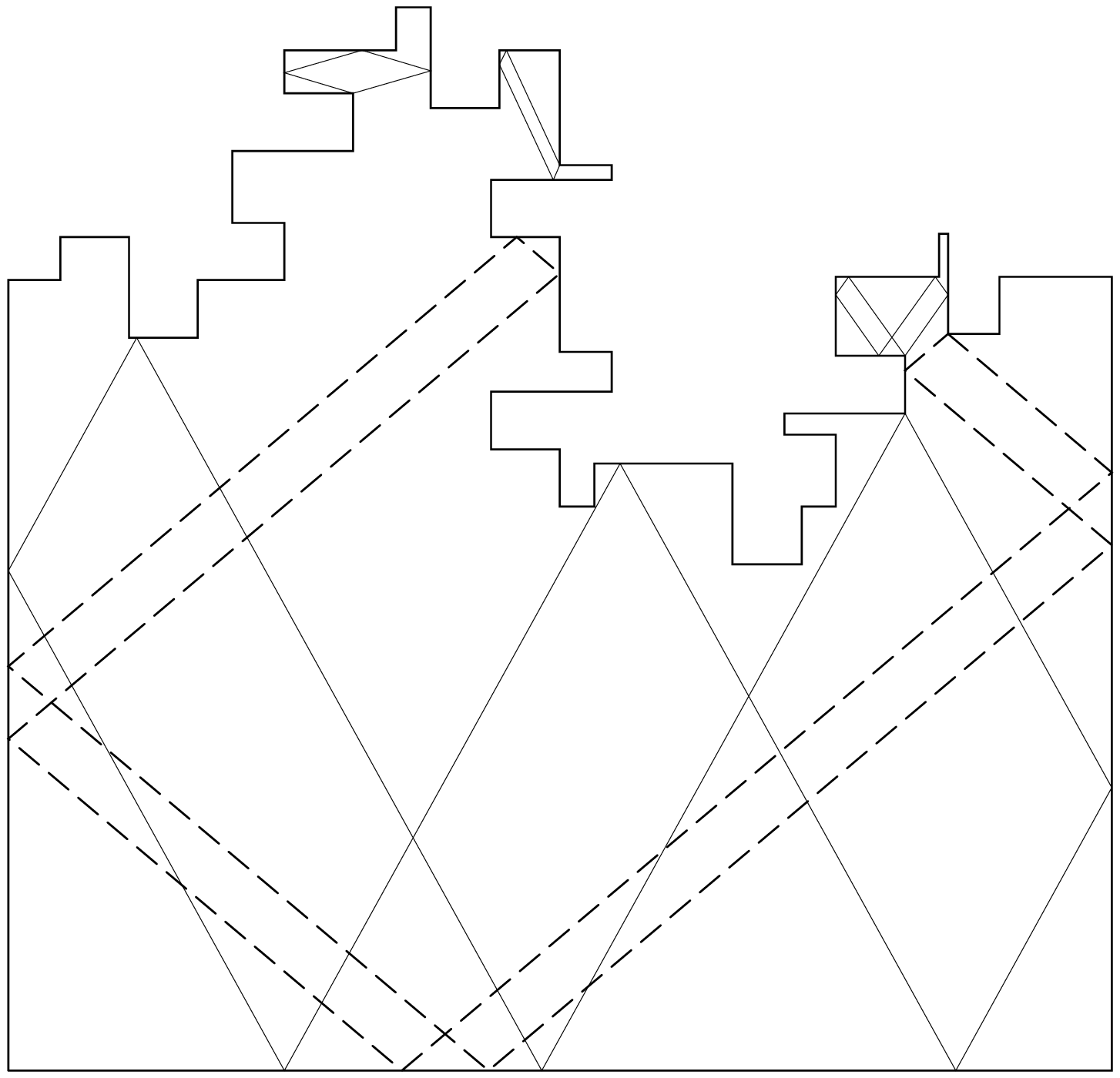}}
\def\epsfsize#1#2{0.52#1}
\put(19,40){\makebox(1,1){{$\displaystyle (\mathrm{a})$}}}
\put(45,38){\makebox(1,1){{$\displaystyle (\mathrm{b})$}}}
\put(5,27){\makebox(1,1){{$\displaystyle (\mathrm{c})$}}}
\put(40,20){\makebox(1,1){{$\displaystyle (\mathrm{d})$}}}
\put(3,37){\line(1,0){12}}
\put(18,37){\vector(1,0){3}}
\put(23,37){\line(1,0){3}}
\put(26,40){\line(1,0){3}}
\put(29,34){\line(1,0){3}}
\put(32,37){\line(1,0){3}}
\put(26,37){\line(0,1){3}}
\put(29,34){\line(0,1){6}}
\put(32,34){\line(0,1){3}}

\put(37,30){\makebox(1,1){{$\displaystyle \mathrm{Y_1}$}}}
\put(68,31){\makebox(1,1){{$\displaystyle \mathrm{Y_2}$}}}
\put(60,23){\makebox(1,1){{$\displaystyle \mathrm{X_1}$}}}
\put(55,44){\makebox(1,1){{$\displaystyle \mathrm{X_2}$}}}

\end{picture}
\caption[]{\small Geometry and periodic orbits: (a) The fractal generator.
(b) The L-shaped system of $g=2$. (c) Fractal drum of $\nu=2$ with the 
generator applied to the upper border and $g=28$. (d) Fractal drum of $\nu=1$ 
with the generator applied to the upper border and $g=4$. The 
segments of the drums are slightly deformed in order to avoid degeneracies of 
the orbit lengths. Some selected periodic orbits are shown for each geometry 
and the beam-splitting property of the salient corners and the different 
segments $X_1$, $X_2$, $Y_1$ and $Y_2$ are demonstrated in (b). 
}
\label{bi:geo}
\end{figure}}

\section{Results and Conclusions}

We consider the systems shown in Fig.~\ref{bi:geo}, i.e. two fractal drums
of $\nu=1$ and $2$, and the simple
two-step system that is shown for comparison.
First, we determined $\Delta_3(L)$ by Eq.~(\ref{delta3}) from the eigenvalues 
calculated numerically by the Lanczos algorithm. 
Second, we calculated the periodic orbits by the hypothetical 
orbit method \cite{biswas3,mellenthin04} that uses the fact that the systems
are constituted by segments of lengths $X_i$ and $Y_j$ in
$x$- and $y$-direction, respectively (compare 
Fig.~\ref{bi:geo}(b)). All periodic orbits have to pass through the segments an
integer number of times. Therefore, only certain discrete angles
$\varphi$ (between trajectory and boundary) and lengths $\ell_\varphi$ are
possible, where 
$\tan \varphi=\sum_i n_i Y_i/\sum_j m_j X_j$, 
$\ell_\varphi=2 \left[(\sum_i n_i Y_i)^2+(\sum_j m_j X_j)^2\right]^{1/2}$, with 
positive integers $n_i$ and $m_j$. This restricts the number of possible orbits
to certain pairs $(\varphi(n_i,m_i),\ell(n_i,m_i))$, which have to be tested.
Due to the shielding role of the corners, not all hypothetical orbits really occur 
in each system. Therefore, the hypothetical orbit method checks
which trajectories actually return to their starting point within the correct length 
$\ell_\varphi$ of the trajectory. Unfortunately, the number of hypothetical orbits
increases with the number of segments by a power-law and the method becomes very
time-consuming for systems with many corners.

As a test for the orbit families we use a sum rule for the 
number $N(\ell)$ of orbits with lengths smaller than $\ell$ (proliferation rate)
\cite{biswas1,biswas2}, 
\beq\label{proliferat}     
N(\ell) = \fr{\pi b_0\ell^2}{\lan a(\ell)\ran},   \qquad
b_0\ell\approx S(\ell) = \fr{1}{2\pi}\sum_{i,\ell_i<\ell} \fr{a_i}{\ell_i} 
\eeq             
where $\lan a(\ell)\ran=\sum_{i,\ell_i<\ell} a_i/\sum_{i,\ell_i<\ell} 1$ is the average
area in phase space, occupied by the orbits with lengths smaller $\ell$ and $b_0$ is 
a constant, depending slightly on the details of the system. 
On systems with $g\le 3$, it has been found that 
$b_0=1/4$ for integrable systems and slightly larger for pseudointegrable systems, whereas 
$\lan a(\ell)\ran$ is saturating to
$\lan a(\ell)\ran\approx 4A$ in integrable systems and to a much smaller value for 
pseudointegrable systems \cite{biswas2,mellenthin04}.
In Fig.~\ref{bi:area}(a), we
plot the normalized average area $\lan a(\ell)\ran/(4A)$ versus $\ell$ for 
the systems of Fig.~\ref{bi:geo}(b-d) with $g=2$, $4$ and $28$. 
The decrease of $\lan a(\ell)\ran$ with $g$ is very drastic, which shows clearly
the beam-splitting property of the salient corners. Each
orbit that is disturbed by a salient corner splits into  
two different families, each of them covering a smaller area than without this
corner. Therefore, $\lan a(\ell)\ran$ 
decreases with $g$, whereas $N(\ell)$ increases.
In Fig.~\ref{bi:area}(b), we plot $S(\ell)$ for the same systems and determine the constant
$b_0$ by the slopes. We can see that $b_0$ is very close to $0.25$ (the value of integrable systems)
for the system with $g=2$ and $4$ and considerably larger for $g=28$. 

With the values of $b_0$ and $\lan a\ran$, we can test the proliferation law $N(\ell)$ of
Eq.~(\ref{proliferat}) for each system (see Fig.~\ref{bi:proliferat}(a)).
The lines show the theoretical curves, whereas the symbols indicate the number $N$
of orbits found by our calculations. They agree very 
well for $g=2$ and $4$, whereas deviations from the quadratic shape of
$N(\ell)$ occur for $g=28$, showing that many comparably small orbits are
still missing in the spectrum of the $\nu=2$-system. Due to the three different length 
scales of
this drum and to the large number of orbit families, it is very time-consuming to
find a sufficient number of them. However, we can already see by the present data that
$N(\ell)$ increases drastically with $g$.

\unitlength 1.50mm
\vspace*{0mm}
\begin{figure}
\begin{picture}(0,30)
\def\epsfsize#1#2{0.3#1}
\put(5,0){\epsfbox{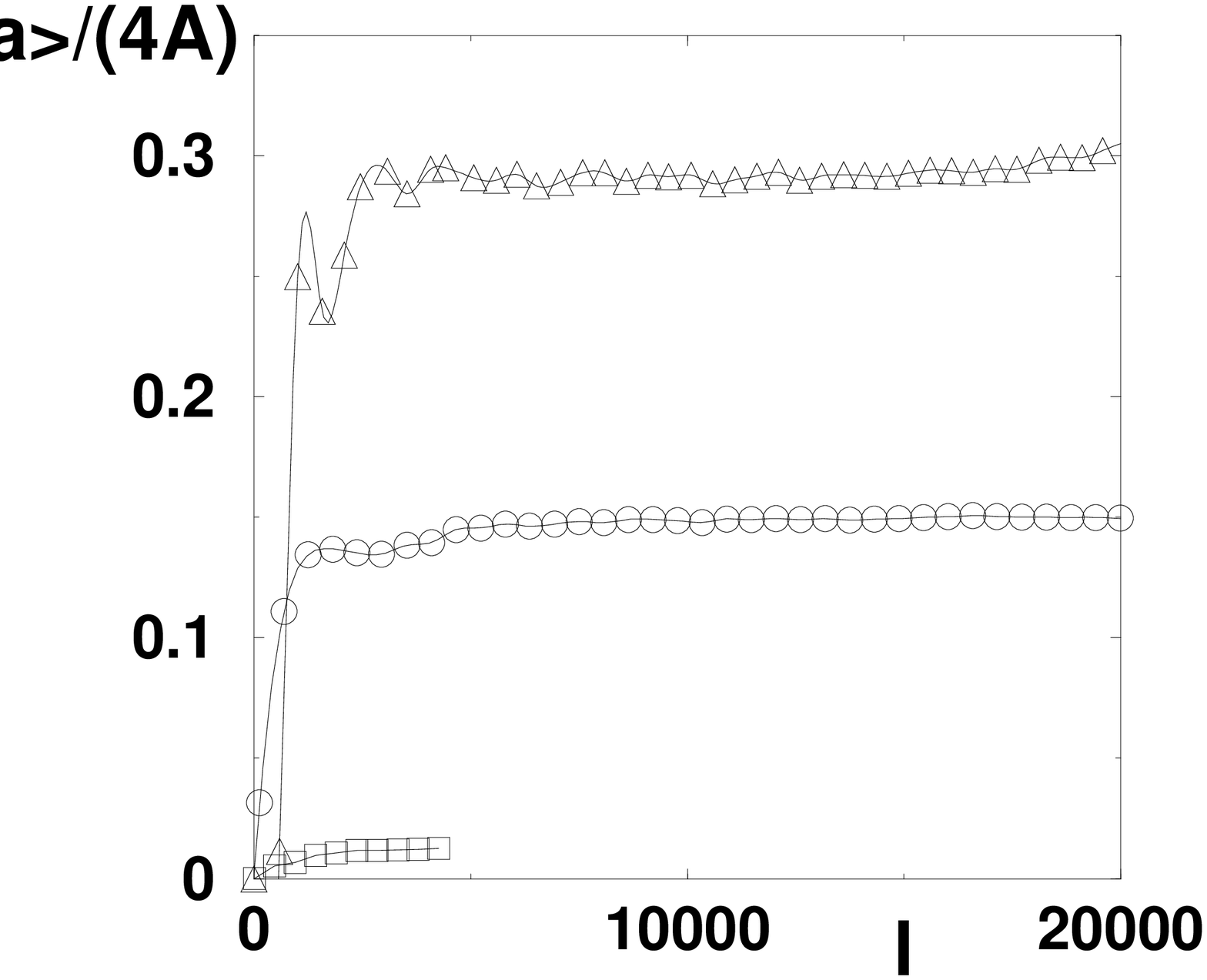}}
\put(50,-2){\epsfbox{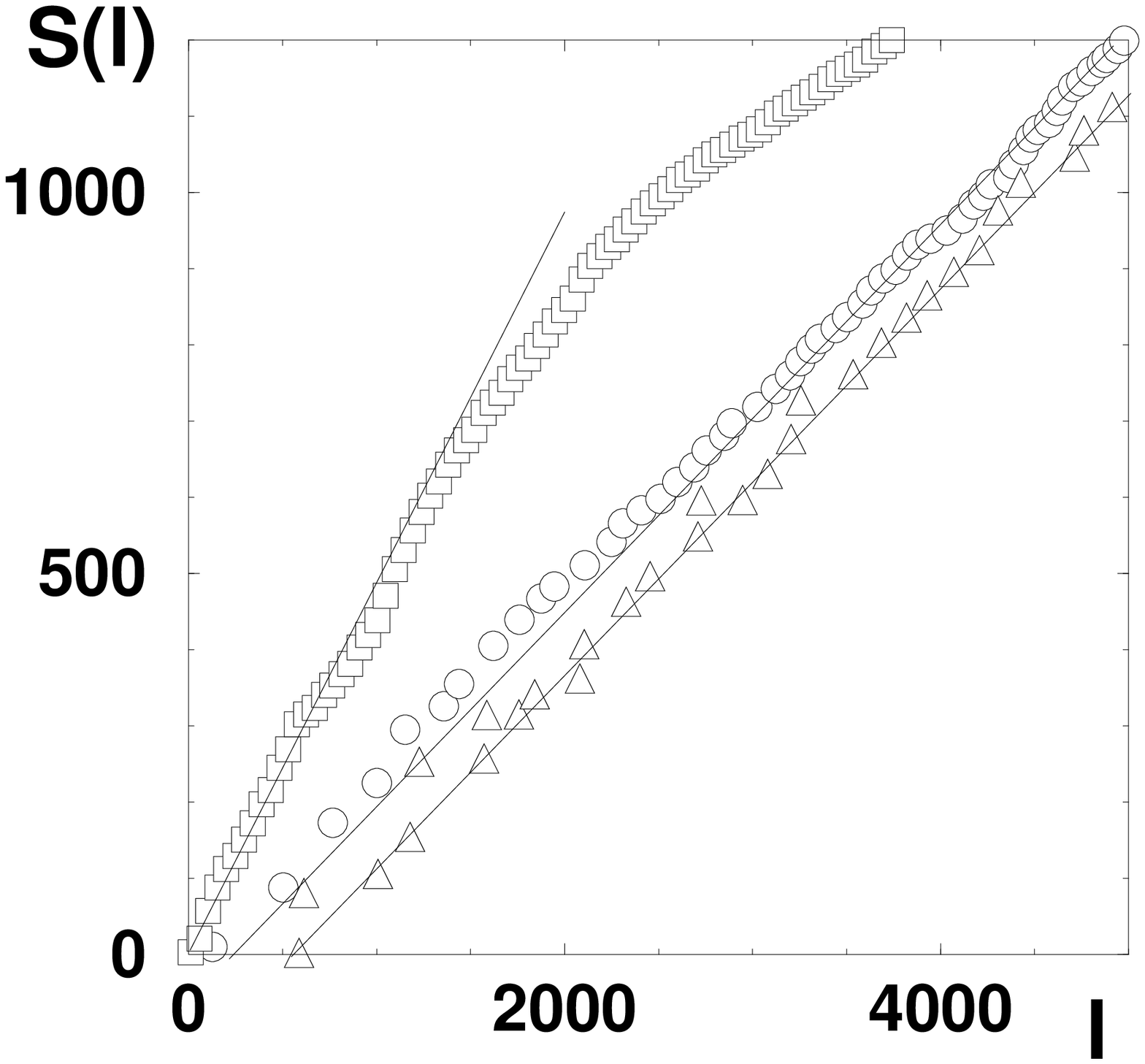}}
\put(30,6){\makebox(1,1){{\large \bf (a)}}}
\put(76,5){\makebox(1,1){{\large \bf (b) }}}
\end{picture}
\caption[]{\small (a) The normalized average areas $\lan a\ran/(4A)$ of the 
periodic orbit families, 
with the area $A$ of the systems is plotted versus the orbit length $\ell$ for 
the systems of Fig.~\ref{bi:geo} with $g=2$ (triangles), $g=4$ (circles) and 
$g=28$ (squares). 
(b) $S(\ell)$ for the same systems is plotted versus $\ell$.
}
\label{bi:area}
\end{figure}

\unitlength 1.50mm
\vspace*{0mm}
\begin{figure}
\begin{picture}(80,30)
\def\epsfsize#1#2{0.3#1}
\put(5,-2){\epsfbox{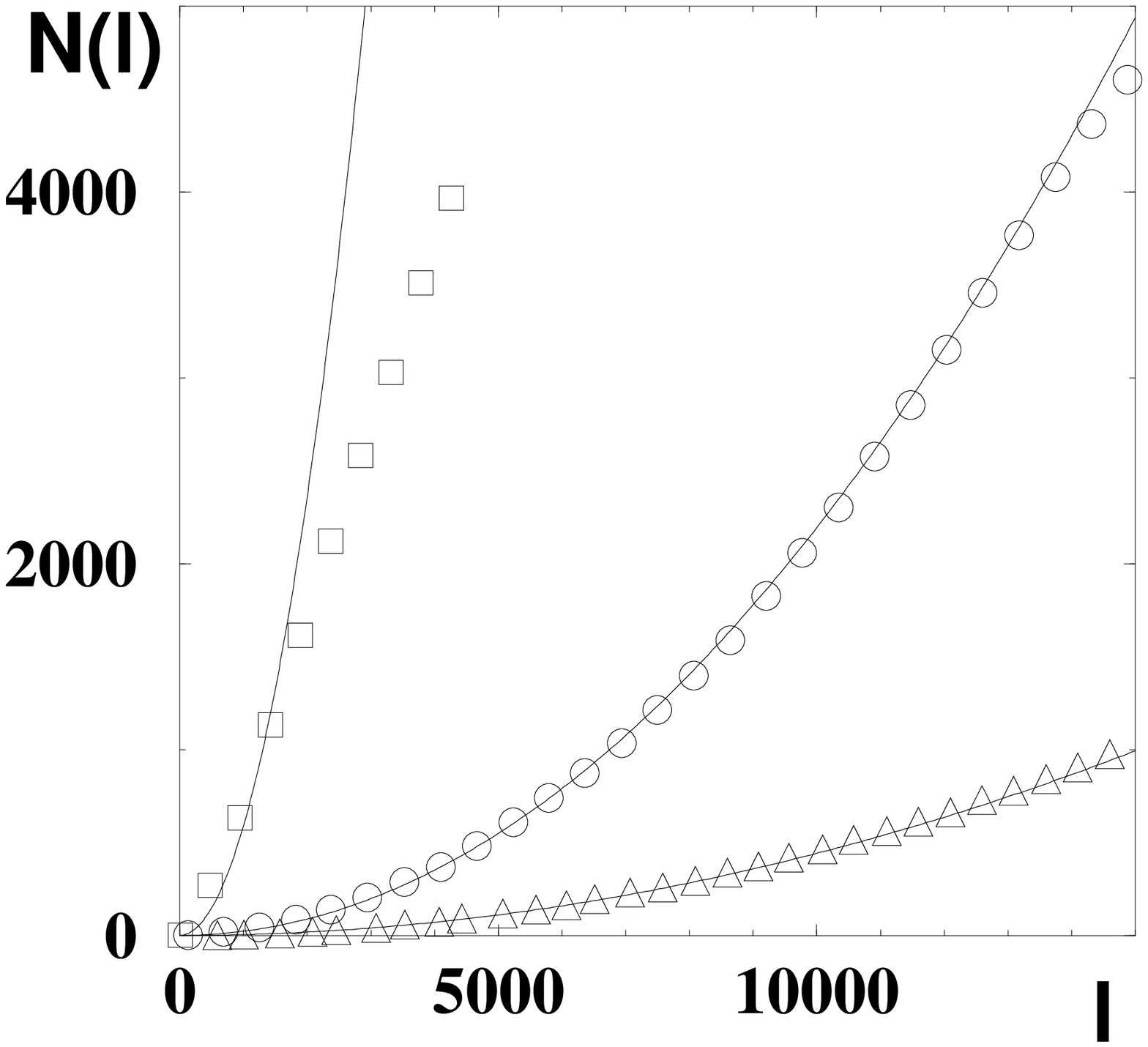}}
\put(50,0){\epsfbox{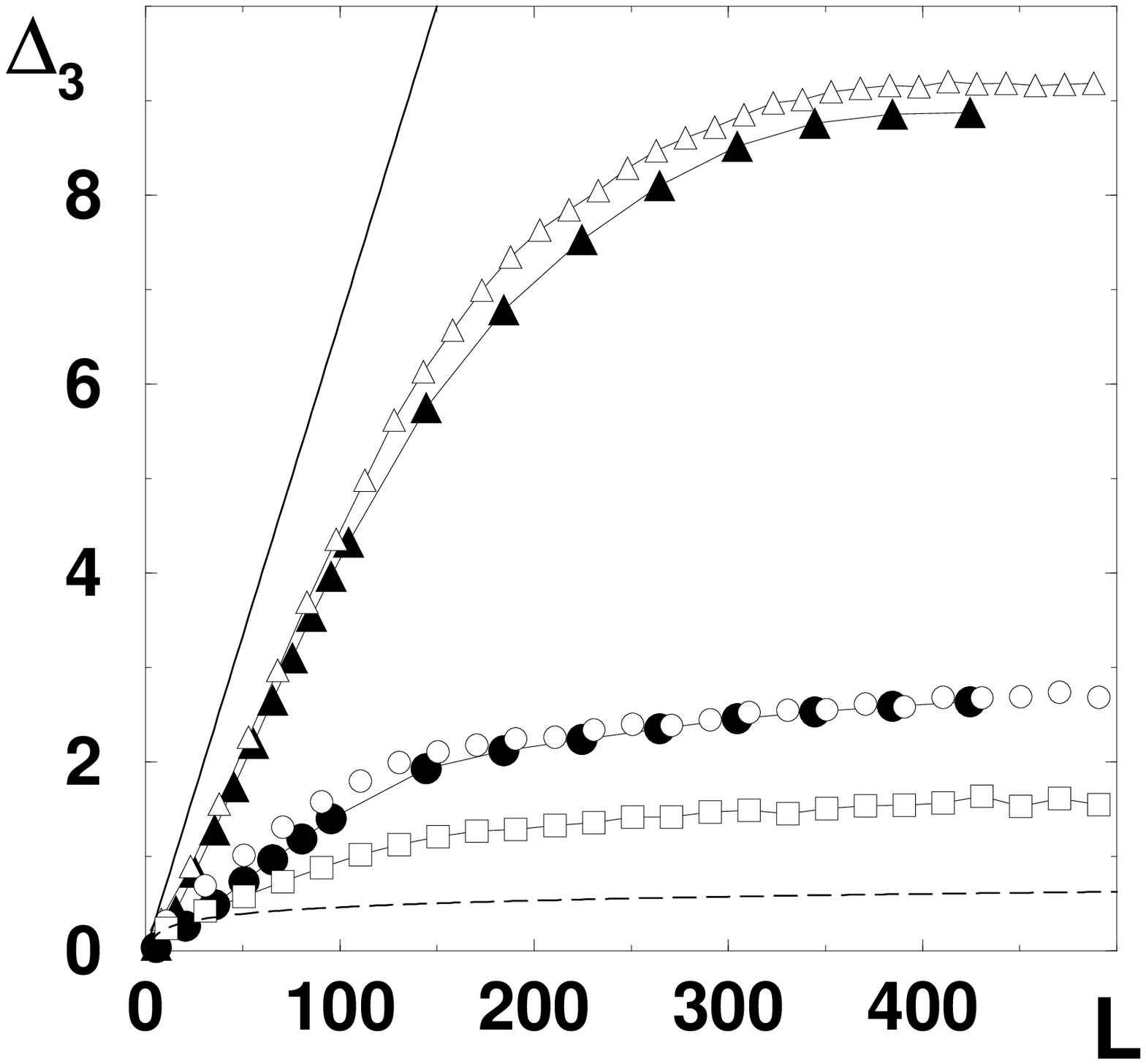}}
\put(22,25){\makebox(1,1){{\large \bf (a) }}}
\put(55,25){\makebox(1,1){{\large \bf (b) }}}
\end{picture}
\caption[]{\small (a) The proliferation rate $N(\ell)$ is plotted versus $\ell$.
The lines show the theoretical curves, whereas the symbols indicate the number
of orbits that have been found by our calculations.
(b) $\Delta_3(L)$ is plotted versus $L$. Filled symbols refer to periodic-orbit 
calculations (Eq.~(\ref{biswaseq})) and open symbols to calculations from the eigenvalues 
(Eq.~(\ref{delta3})). 
(Same symbols as in Fig.~\ref{bi:area}.)
The expected curves for integrable and chaotic systems are
indicated by a solid and a dashed line, respectively.
}
\label{bi:proliferat}
\end{figure}

We finally calculate $\Delta_3(L)$ (i) from Eq.~(\ref{delta3}) using the eigenvalues
and (ii) -- for the systems of $g=2$ and $4$, where $N(\ell)$
has been reasonably well approximated by the calculated orbits -- by periodic orbit theory.
In both cases, the data from the eigenvalue and from the periodic orbit calculations
agree very well. This indicates that higher-order terms,
arising e.g. from diffractive orbits (additional to the
periodic orbits) seem to be small (at most in the order of
magnitude of the error bars arising from the Lanczos algorithm and the discretization
of the system). It will be very
interesting to see, if this stays true for the fractal drum of $\nu=2$, where many more
corners and therefore diffractive terms are present.
We also see in Fig.~\ref{bi:proliferat}(b) that the curves decrease rapidly with
increasing $g$, thereby coming closer to the Wigner curve for chaotic systems. It
will be very interesting to see, if drums of higher fractal generations $\nu$ can come 
arbitrarily close
to chaotic systems and in which way lengths and areas of the periodic orbits change with
$g\to\infty$.

\end{document}